\documentstyle[aps,epsf]{revtex}
\begin{document}
\draft

\title{Two point-contact interferometer for quantum Hall systems}

\author{C. de C. Chamon}
\address{Department of Physics, Massachusetts
Institute of Technology, Cambridge, MA 02139}
\author{D. E. Freed}
\address{The Mary Ingraham Bunting Institute, Radcliffe
Research and Study Center, Harvard University, Cambridge, MA 02138}
\author{S. A. Kivelson}
\address{Department of Physics, University of California at Los Angeles,
Los Angeles, California 90095}
\author{S. L. Sondhi}
\address{Department of Physics, Princeton University,
Princeton, NJ 08544}
\author{X. G. Wen}
\address{Department of Physics, Massachusetts
Institute of Technology, Cambridge, MA 02139}
\maketitle
\begin{abstract}
We propose a device, consisting of a Hall bar with two weak barriers,
that can be used to study quantum interference effects in a strongly
correlated system. We show how the device provides a way of measuring
the fractional charge and fractional statistics of quasiparticles in
the quantum Hall effect through an anomalous Aharanov-Bohm period. We
discuss how this disentangling of the charge and statistics can be
accomplished by measurements at fixed filling factor and at fixed
density. We also discuss a another type of interference effect that
occurs in the nonlinear regime as the source-drain voltage is varied.
The period of these oscillations can also be used to measure the
fractional charge, and details of the oscillations patterns, in
particular the position of the nodes, can be used to distinguish
between Fermi-liquid and Luttinger-liquid behavior. We illustrate
these ideas by computing the conductance of the device in the
framework of edge state theory and use it to estimate parameters for
the experimental realization of this device.
\end{abstract} 
\pacs{PACS: 73.23.-b, 71.10.Pm, 73.40.Hm, 73.40.Gk}


\section{Introduction}
\label{intro}
A considerable amount of work on electronic systems in recent years
has focused on two distinct sets of problems: the effects of
quantum interference on the behavior of mesoscopic systems and those
of strong correlations in low dimensional systems. Most of the
canonical work on the former \cite{Altshuler-Lee-Webb} has involved
single electron physics while work on strongly correlated electrons
has, by definition, been concerned with the effects of inter-electron
interactions. Recent advances in semiconductor device fabrication have
led to a convergence of these streams of work, in that it is now
possible to conduct experiments that test quantum interference in
strongly interacting systems of electrons.

The work reported here takes advantage of this convergence. Our
principal motivation is the physics of the fractional quantum Hall
(FQH) states, which exhibit some of the most striking effects of
strong electronic correlations. These are perhaps most evident in the
unusual quantum numbers of FQH quasiparticles: they are fractionally
charged \cite{Laughlin}, obey fractional statistics
\cite{Halperin,Arovas} and couple to curvature with a fractional spin
\cite{Einarsson,XGW&Zee,DHLee}. These correlations also lead to a
novel dynamics at the edges of FQH systems which is that of one
dimensional chiral Luttinger liquids \cite{XGWcll}.

Our chief purpose in this paper is to describe and analyze a device,
the two point-contact interferometer, that would allow direct
observation of the fractional charge {\em and} statistics of the
quasiparticles as well as allow tests of the chiral Luttinger liquid
behavior of the edges; the former function is largely independent of
and more robust than the latter.  This device has remarkably rich
interferometric possibilties; it exhibits conductance oscillations
with varying magnetic field and also with varying amplitude of the
voltage across it.

The paper is organized as follows. In the balance of the introduction
we discuss some related work. In Section \ref{secQual} we describe the
interferometer and give a qualitative discussion of its physics in a
largely model independent way.  In Section \ref{secModel} we introduce
a model, defined in terms of edge state theory, that allows explicit
calculations of the conductance of the device. We treat the model
within perturbation theory and solve for the transmission current
which displays oscillations with both magnetic field and voltage. In
Section \ref{secg=1} we consider the exactly solvable case in which
the edge states are chiral Fermi liquids. This is the case for edge
states of an integer filling factor state ($\nu=1$), and helps provide
an intuitive understanding of the voltage interference patterns for
general $\nu$.  Finite temperature effects are treated in Section
\ref{secfiniteT}, where we show how the oscillations are washed out as
the temperature is raised.  In Section \ref{estimates}, we give
numerical estimates for the sizes of the parameters at which the
Aharonov-Bohm and voltage oscillations occur.  The appendix contains
the details of the perturbative calculation of the transmission
current.

\subsection{Related Work}

The direct observation of fractional quantum numbers in FQH systems
has long been of interest. Pokrovsky and Kivelson \cite{Kivelson1}
suggested ways in which the fractional charge could be detected
through a fractional Aharanov-Bohm period. This has been elaborated
further in the work of Kivelson \cite{Kivelson2} and of Pokrovsky and
Pryadko \cite{Pokrovsky}. Strong evidence of such oscillations was
found in the experiment of Simmons et al \cite{Simmons}, who measured
conductance oscillations on the edges of various QH plateaux. However
it has not been clear what microscopic details of the transport in
that region led to these oscillations. It is our current belief that
most likely their sample fortuitously realized a version of the
interferometer discussed here. Recent experiments involving tunneling
across an antidot in a FQH sample \cite{Goldman,Mace}, have provided
convincing evidence for a fractional local charge \cite{Goldhaber}
that couples to the electrochemical potential. Finally, Kane and
Fisher \cite{Kane&Fisher2} and Chamon, Freed and Wen
\cite{XGW2,CFW,CFW2} have suggested that measurements of the noise for
tunneling currents between the edges of a FQH system, i.e.  in a
single point-contact device, could be used to detect the fractional
charge; calculations of the zero frequency noise for the exactly
integrable model have been carried out by Fendley et al.
\cite{Fendley3}.

The detection of fractional statistics remains unaddressed by
experiments to date. The theoretical basis for such a detection was
first discussed by Kivelson \cite{Kivelson2} and an intriguing
proposal involving flux periodicities for hierarchy droplets inside
FQH systems has been proposed by Jain, Thouless and Kivelson\cite{Jain1}.
The observation of the fractional spin seems quite difficult and still
awaits a concrete scenario for an experiment.

Finally, our concrete discussion of the physics of the interferometer
in the framework of edge state theory \cite{XGWreview} expands the
scope of the work on tunneling in chiral Luttinger liquids by Wen
\cite{XGW2}, and the work on resonant impurity tunneling in Luttinger
liquids by Kane and Fisher \cite{Kane&Fisher1}. A very recent paper by
Geller {\it et al} \cite{Geller} discusses resonant tunneling through
anti-dots and although theirs is a distinct geometry, and they
consider only electron tunneling, their treatment is similar to ours.

\section{Device Description and Qualitative Discussion}
\label{secQual}

\begin{figure}
\vspace{.8cm}
\hspace{1.15 in}
\epsfxsize=4.5in
\epsfbox{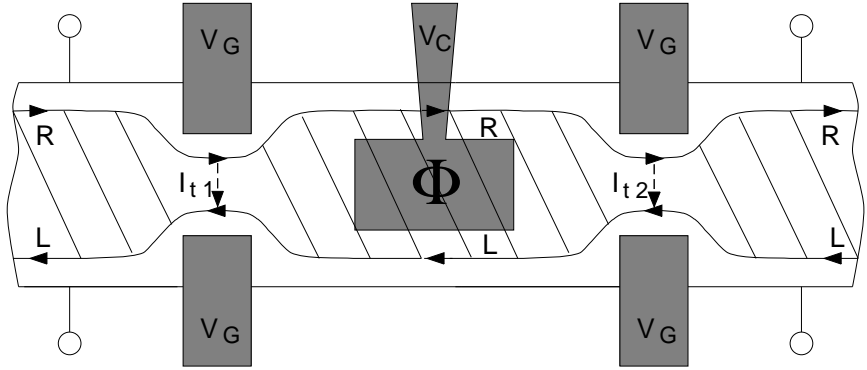}
\vspace{.8cm}
\caption{Two point-contact interferometer. Two gates are
placed a distance $a$ apart. The gate voltages are adjusted so as to
bring the edges of a FQH state with filling fraction $\nu$ close
together, but not pinch the constriction. In this way, quasiparticles
carrying fractional charge and statistics can tunnel from one edge to
the other. A magnetic flux $\Phi$ can be inserted in the region
between the point-contacts that is bounded by the edge states. A
central gate allows the charge in the region to be selectively
depleted.  The transmitted current (the Hall current $I_H=\nu e^2/h$
minus the tunneling current $I^1_t+I_t^2$) oscillates as a function of
the inserted flux, the voltage difference between the edges and the
voltage of the central gate. An overall back gate on the device allows
magnetic field sweeps at constant filling factor.}
\label{fig1}
\end{figure}

The device we are proposing consists of three components, as
illustrated in Figure \ref{fig1}. The first and primary component is a
narrow quantum Hall bar with two tunable constrictions, or
point-contacts, whose separation is less than a phase coherence length
at low temperatures.  The second component is a back gate that allows
the electron density in the Hall bar to be varied uniformly. The third
component is another gate ({\it e.g.} an air bridge) that would allow
the center of the region defined by the two point-contacts to be
selectively depleted by the application of a voltage. Estimates for
the dimensions of the device, which seem entirely feasible with
existing fabrication techniques, are discussed in Section
\ref{estimates}; here we note that these require that the point
contacts be a few microns apart for operating temperatures of 100mK
and below and that the electron gas be about 1000$\AA$ or less from
the point-contacts and the central gate. The back gate is not required 
to be particularly close to the electron gas.

The physics of this device is that of a quantum version of the
Fabry-Perot interferometer\cite{Yariv}. However, our interferometer
differs from the standard non-interacting one in that the scattering
particles cannot be assumed to be independent because of the strong
correlations. At values of the magnetic field where the electrons in
the bulk of the device are deep in a QH phase, the low energy
excitations, or quasiparticles, lie on the edges of the bar. At the
constrictions, they can tunnel from one edge to another and the
resulting tunneling current will cause the Hall conductance to deviate
from its quantized value, as in the case for a single point-contact.
However, having two tunneling sites results in phase sensitivity of
the tunneling current; tunneling events taking place at one of the
contacts will interfere with those occuring at the other.

These interference effects can be modulated in three distinct ways:
The first involves changing the magnetic field and leads to what we shall
call Aharanov-Bohm (AB) oscillations for obvious reasons. The second
involves changing the number of quasiparticles enclosed by the
interfering orbitals and leads to (fractional) statistical
oscillations. The third involves varying the source-drain voltage and 
leads to what we shall call a novel set of ``Fabry-Perot''
oscillations. In the following we will show how these various effects
can be disentangled to provide a means of measuring both the
fractional charge and fractional statistics of the quasiparticles as
well as to probe the non-Fermi (Luttinger) liquid behavior of the
edges of FQH states.

We would
like to emphasize that we will always be interested in the limit where
the barriers are weak, so that the constrictions are far from being
pinched off, as in Figure \ref{fig1}. (The opposite limit, where the
two point-contacts are near pinch off, is similar to tunneling through
a quantum dot, except that for our geometry the central island would
be larger. It can be analyzed using methods similar to those in this
paper, but we will not address it here.) The chief advantage of this
restriction is that we stay far from the regime near pinch off where,
experimentally, poorly understood resonances arise already for a
single point-contact \cite{Webb}. Consequently, we expect that the only
resonances are the ones explicitly created by the two point-contact 
geometry and the resulting energy and field scales for these are set by 
the parameters of the device and can be chosen to lie in an observable
range. (Given the lack of understanding of the near pinch-off resonances,
it is hard to say theoretically whether they are absent for weak barriers;
however, experiments involving antidot resonances \cite{Goldman,Mace}
do strongly suggest this.) 
Further, it becomes possible to probe the internal structure of
the resonance, i.e.  its intermediate energy behavior, without detailed
microscopic knowledge. Another restriction on our analysis is that we
consider only the primary Hall states, i.e. $\nu=1/m$ with $m$ odd,
where there is only one branch of edge excitations and life is
somewhat simpler; the extension to their descendant states does not
pose any conceptual problems.

\subsection{Aharonov-Bohm and fractional statistics oscillations}

We will first discuss the interference effects which occur when the
magnetic field is varied. Consider the transmission amplitude for
quasiparticles propagating along the right edge. As they can tunnel to
the left edge at the two constrictions in Fig. \ref{fig1}, the
amplitude will involve a sum over paths that encircle the area $A$
enclosed by the edges and the constrictions any number of times. As a
result, they pick up an AB phase proportional to the flux
$\Phi$ through this area.  Naively, this phase is given by $2\pi e^* B
A /(hc)$, where $e^*$ is the charge of the quasiparticle. It is
convenient to define an effective flux quantum by $\Phi^* ={e\over
e^*}\Phi_0$, where the usual flux quantum is given by $\Phi_0= hc/e$.
Then, in terms of this effective flux quantum, we would expect that as
the magnetic flux is varied, the current and other properties of the
system would undergo oscillations with period $\Delta B^* = \Phi^*/A$,
and thus measurements of these oscillations would provide a means of
measuring the fractional charge.

However, this conclusion is too naive. The quasiparticles derive their
properties from the parent liquid which is the relevant ``vacuum'' and
{\em only} when the vacuum is invariant can we expect to use arguments 
based solely on their AB phases. Indeed, if the extra flux added is
dynamically localized in the interior of the fluid this would effectively
create a multiply connected geometry where gauge invariance for the 
constituent electrons implies a
flux periodicity of $\Delta B = \Phi_0 / A$. As this is smaller
than the quasiparticle period $\Delta B^*$, this would exclude
oscillations with the latter periodicity. We should emphasize that this
is a {\em dynamical} possibility, there are no general, non-trivial 
consequences of gauge invariance for the geometry at issue here. At
any rate, it is clear that we need to be careful about considering
changes in the bulk of the fluid as the flux is varied. To this end we 
distinguish between two cases.

\noindent
{\it i}){\em Field sweeps at fixed particle number:} In this case, as
we just observed, we expect to observe conductance oscillations with
period $\Phi_0/A$. This can happen in one of two ways depending upon
the detailed electrostatics of the central region. If, as the magnetic
field is raised, the QH fluid in the central region shrinks uniformly
in order to keep the filling fraction constant, then its area
decreases by just the right amount to leave the flux through the whole
droplet unchanged. Consequently, there are no oscillations as the
magnetic field is varied and the period is trivially $\Phi_0/A$. If,
instead, the electrostatics prefers otherwise and the self-consistent
potential has a maximum in the interior, quasiholes will be created
there as the magnetic field is increased---one quasihole for each flux
quantum. The phase picked up by a quasiparticle encircling the central
region is now the sum of the fractional AB phase and the phase
$-\theta^* = -2\pi/m$ due to its (fractional) statistical interaction
with the central quasihole and precisely restores the periodicity to
$\Delta B$.

\noindent
{\it ii}){\em Field sweeps at fixed filling factor:} One obtains quite
different results if the field is swept at fixed filling factor. In
this case the quasiparticles see an invariant vacuum (in other words,
the electron density is changed so that quasiholes are not formed in
the bulk of the droplet). The resulting periodicity is then $\Delta
B^*$. The observation of conductance oscillations with such a
fractional AB period would constitute a measurement of a fractional AB
charge for the quasiparticles. Experimentally, keeping the filling
factor constant requires changing the number of particles along with
the field which is why the device requires a back gate. Also,
preventing the formation of quasiholes in the central region requires
that net fractional charge be added to the this area, which is only
possible if the contacts are not pinched off.

Having discussed the observation of fractional charge we now turn to
the observation of fractional statistics. We first note that the
comparison between the periodicity $\Delta B$ when the number density
$n$ of electrons is held constant and the periodicity $\Delta B^*$
when the filling fraction $\nu$ is held constant, implicitly verifies
the fractional statistics of the quasiparticles and quasiholes,
because in one case the period is due to the combined effects of the
Aharonov-Bohm phase and fractional statistics, and in the second it is
due only to the Aharonov-Bohm phase.

To more directly see the effect of the fractional statistics, we need
to consider oscillations that arise from having varying numbers of
quasiparticles in the central region.  If $N$ quasiholes are present,
then the interference phase is modified to $2\pi (B A/\Phi^* -N/m)$. It
is clear then that it is neccessary to add $m$ quasiparticles before
the interference condition is restored. To this end one can imagine
using the central gate to deplete the central region in steps of
charge $1/m$ which would then lead to conductance oscillations with a
period of $m$ steps. A better strategy, which requires less control
over the electrostatics, is to create some unknown number of
quasiholes in the central region and then look for a shifted
fractional AB pattern at fixed filling factor, as in the charge
measurement. Except when an integer multiple of $m$ quasiholes are
present, there will be a shift in this pattern and the observation of
$m-1$ distinct shifts will be a direct signature of the statistical
interaction between the quasiparticles.  For example, at $\nu=1/3$,
there will be two shifted patterns with shifts of $(1/3)\Delta B^*$
and $(2/3)\Delta B^*$.
 
\subsection{Fabry-Perot oscillations due to finite
source-drain voltages}

The third modulation of the interference appears when varying the
source-drain voltage $V$, and again leads to oscillations in the
conductance. The origin of these oscillations is most transparent for
$\nu=1$, where single-particle considerations suffice; this is
detailed in Section \ref{secg=1}. In brief, the conductance at finite
$V$ is determined by the transmission of electrons in a range of
energies $\Delta E = e V$ while the transmission itself oscillates
with energy.  Consequently, the integrated transmission, and hence the
conductance itself, oscillates with the source-drain voltage, albeit
with an envelope that decreases as $1/V$.

An interesting perspective is afforded by thinking in analogy to the
classical wave analysis of the Fabry-Perot interferometer, {\it i.e.}
a device with two parallel partially transmitting barriers; as we
shall see in the edge state analysis of the device, it is conceptually
exact and allows a unified treatment of the fractional fillings as
well.  Evidently, the combination $\omega_{\rm osc}=2\pi / \tau$, the
inverse time for edge waves (and hence the quasiparticles) to travel
from one point-contact to the other, is the characteristic frequency
of the device and will set the scale for the oscillations in its
transmission due to multiple reflections within it. The source-drain
voltage defines a second frequency scale, the Josephson frequency
$\omega_J = e^*V/\hbar$, where $e^*$ is the fractional charge of the
quasiparticles living on the edges; this is the bandwidth of the waves
incident on the interferometer. It follows then that the transmission
will be a function of the ratio $\omega_J/\omega_{\rm osc}$ as well as
of the reflection/transmission coefficients at the point-contacts
which are determined by the quasiparticle tunneling amplitudes at
them.

Because the right-moving and left-moving waves lie on the edge of a QH
droplet, there is a second perspective that is very illuminating.
(This picture, however, is not as general as the actual calculation of
the voltage oscillations, since the calculation does not explicitly
require the left- and right-moving edges to enclose any amount of
flux; it would also be valid for a 1D wire if the modes were to have
differing chemical potentials.) In this picture, the Aharonov-Bohm
phase is responsible for the oscillations in the transmission as the
energy is varied because the different energies lead to different
areas enclosed by the interfering orbits.  More specifically, an
increase in energy of $\delta E$ corresponds to a change in the
momentum of the quasiparticles given by $\delta k =\nu \delta E/(\hbar
v)$, where $v$ is the velocity of the edge modes \cite{footnote2}.
The momentum at the edge is related to the area of the droplet $A$ and
its perimeter $2a$ by $k = \nu A/(l^2 2a)$, where $l$ is the magnetic
length, so that an increase of $\delta E$ in energy results in an
increase in area of $\delta A = 2 l^2 a \delta E/(\hbar\nu v)$.  The
extra flux enclosed then gives a change in phase of $2\pi\delta
\Phi/\Phi_0^* = 2a \delta E/\hbar v = 4\pi\delta E /\hbar\omega_{\rm osc}$. As
a result, as the energy is varied, the transmission oscillates with a
period set by $\omega_{\rm osc}$
\cite{footnote3}.

Three conclusions follow from this description. First, we recover the
result that the net transmission has a component that oscillates with
the source-drain voltage. Second, we find that there are two distinct
regimes as a function of $\omega_J/\omega_{\rm osc}$. For small values
of this ratio the device is probed at low frequencies and hence at a
long length scale where the coherence between the two barriers is
important. In this regime the phase difference between the
reflection/transmission coeffecients at the barriers governs the
transmission. Because this phase difference is determined by the AB
phase, the AB oscillations discussed in the previous section will
occur.  At large values of the ratio, the barriers enter independently
and the AB oscillations are washed out.  The third and most
interesting conclusion is that {\em even} for intermediate values of
the ratio there are special points where the AB oscillations
disappear. Again, the origin of this disappearance is most transparent
for $\nu=1$: the nodes occur whenever the bandwidth is equivalent to a
phase difference of an integer multiple of $2\pi$ across it. In such
cases the phase shift between the barriers is immaterial; essentially,
one is integrating the interference pattern over an integral number of
periods, so the oscillations cancel.

The detailed analysis in Section \ref{secModel}, where we consider the
case of general $\nu$, bears out the same qualitative feature that
there are special values of $V$ where the AB oscillations disappear.
However, the location of these nodes is modified in a very interesting
way which depends sensitively upon the nature of correlations in the
edges. If the edge dynamics are of the Fermi liquid variety, as should
be the case at $\nu=1$, our naive assertions are correct. However, if
the edges are Luttinger liquids then the locations of the nodes are
given by the zeros of Bessel functions which depend on
$\omega_J/\omega_{\rm osc}$, and the type of Bessel function depends
on the Luttinger liquid exponent $g$. The nodes occur for
$\omega_J/\omega_{\rm osc}\approx (n+\xi_g)/2$, $n=0,1,\dots$, with
the $g$ dependent shift $\xi_g=(1+g)/2$. For quasiparticle tunneling,
where $g=\nu$, this shift is not an integer, and can, in principle, be
used to measure $g$! (Given an independent measurement of $v$, this
also allows a measurement of $e^*$.) Thus it follows that the
interferometer can also be used to distinguish between Fermi liquid
(with $g=1$) and Luttinger liquid behavior at the edges of FQH
systems.

In the qualitative discussions of the Aharonov-Bohm and Fabry-Perot
oscillations above, we have considered the zero temperature case for
simplicity. The effects of finite temperature, particularly the 
suppression of quantum interference, are treated quantitatively in 
Section \ref{secfiniteT}.


\section{Tunneling between edge states
in the double point-contact geometry}
\label{secModel}

We will now study the interferometer in the framework of edge states
in the quantum Hall effect, which is better cast in the bosonic
language (for a thorough review, see Ref.
\cite{XGWreview}). Our starting point for studying tunneling in a
double point-contact geometry is the Lagrangian density
\begin{equation}
{\cal L}=\frac{1}{8\pi}[(\partial_t\phi)^2-v^2
(\partial_x\phi)^2]\ -
\ \sum_{i=1,2}\Gamma_i \ e^{-i\omega_J t}
\ \delta(x-x_i)\ e^{i\sqrt{g}\phi(t,x_i)}+H.c.\ \ ,
\label{L}
\end{equation}
with the quantization condition $[\phi(t,x),\partial_t\phi(t,y)]=4\pi
i\delta(x-y)$. (We use this normalization of $\phi$ because it gives
an especially simple expression for the dimensions of the tunneling
operators in terms of $g$. To translate to the conventional
normalization of a 1D electron gas or the sine-Gordon model, $\phi$
must be replaced by $\phi/(2\sqrt{\pi})$.) The voltage difference
between the two edges of the QH liquid determines the Josephson
frequency $\omega_J \equiv e^*V/\hbar$, with $e^*=e$ for electron
tunneling and $e^*=e/m$ for quasiparticle tunneling. In the following
we will set the edge velocity $v=1$.  The two point-contacts are
located at $x_1$ and $x_2$, and their tunneling amplitudes are
$\Gamma_1$ and $\Gamma_2$, respectively \cite{footnote4}.

The first term in the Lagrangian, when considered alone, describes the
dynamics of a free boson field $\phi=\phi_R+\phi_L$, which can be
decomposed into its chiral components $\phi_{R,L}$. These components
describe right and left moving excitations along the edges of FQH
states. Charge density operators can be defined in terms of the
$\phi_{R,L}$ through
$\rho_{R,L}=e\frac{\sqrt{\nu}}{2\pi}\partial_x\phi_{R,L}$.

The second term in the Lagrangian comes from the tunneling between the
edges. The tunneling operators can be written as $\Psi^\dagger_L
\Psi_R$ and $\Psi^\dagger_R \Psi_L$. Right and left moving
electron and quasiparticle operators on the edges of a FQH liquid are
given by $\Psi_{R,L}(t,x)\propto e^{\pm i \sqrt{g} \phi_{R,L} (t,x)}\
e^{\pm 	ik_F x}$, where $g$ is related to the FQH bulk state. For
example, for a Laughlin state with filling fraction $\nu=1/m$ we have
$g=m$ for electrons and $g=1/m$ for quasiparticles carrying fractional
charge $e/m$.  One can verify that $[\rho_{R,L}(t,x)\ ,\
\Psi^\dagger_{R,L}(t,y)]=e\sqrt{\nu g}\
\Psi^\dagger_{R,L}(t,y)\delta (x-y)$, so that indeed the cases
$g=\nu^{-1}$ and $g=\nu$ correspond to the electron ($e^*=e$) and
quasiparticle ($e^*=\nu e$) creation operators, respectively.

The flux $\Phi$ in the area enclosed by the edge branches between the
two point-contacts is taken into account by the phase of the tunneling
amplitudes, $\Gamma_i$, in Equation (\ref{L}) . This phase comes from
the quasiparticle's momentum $k_f$ in the definition of $\psi_{R,L}$.
 From this definition, the tunneling operator $\psi_L^\dagger(x)
\psi_R(x)$ has the phase $e^{2ik_fx}$, where $2k_f$ is equal to the
momentum difference between the right-moving edge and the left-moving
edge. The momentum difference $2k_{fe}$ between electrons on the two
edges is directly related to the perimeter $L$ and area $A$ of the QH
liquid confined between the two point-contacts.  It is given by
$k_{fe} L = 2\pi BA/\Phi_0$. If the distance between the two edges at
each of the point-contacts is much smaller than the distance $a$ along
an edge between the two contacts, then the perimeter can be set equal
to $2a$.  For $\nu = 1/m$, with $m$ an integer, the momentum of the
quasiparticles is then given by $2k_f =
\nu 2 k_{fe} = 2\pi \Phi/(\Phi^* a)$.  Because $\Gamma_1$ is the
amplitude for tunneling at $x = -a/2$, it has the phase $e^{i2\pi
\Phi/(2\Phi^*)}$, and similarly $\Gamma_2$ has the phase $e^{-i2\pi
\Phi/(2\Phi^*)}$.  Thus, the flux can be introduced in Equation (\ref{L})
by taking $\Gamma_{1,2} = {\bar \Gamma}_{1,2}e^{\pm i2\pi
\Phi/(2\Phi^*)}$, where ${\bar \Gamma}_{1,2}$ are couplings
which do not include the phase due to the magnetic flux. Combining
these together, we find that a quasiparticle that circles the area
between the two constrictions once (by tunneling from the left-moving
edge to the right-moving edge at  $x = -a/2$ and tunneling back to
the left-moving edge at $x = a/2$) picks up the amplitude $\Gamma_1^*
\Gamma_2 = {\bar \Gamma}_1^* {\bar \Gamma}_2 e^{-2\pi i\Phi/\Phi^*}$,
so that the phase of the tunneling amplitudes determines the
Aharonov-Bohm phase.

The form of the phases appearing in the amplitudes $\Gamma_1$ and
$\Gamma_2$ can also be viewed as coming from the interaction of the
bosonic edge states $\phi$ with the electric and magnetic fields.  In
particular, the electric and magnetic potentials act as sources which
the field $\phi$ interacts with via its optical charge $\sqrt{\nu} e$
\cite{XGWoptq}. In this way, one can show that the full phase due to
the magnetic flux $\Phi$ and the $N$ quasiholes in the area between
the two constrictions can be accounted for by taking
$\Gamma_1^*\Gamma_2 = {\bar \Gamma}_1^*{\bar \Gamma}_2
e^{-i2\pi(\Phi/\Phi^* -N\nu)}$, when $\nu$ is equal to one over an
integer.  Then, according to Section
\ref{secQual}, the full phase will depend on exactly how the magnetic
field is varied; if the electron number density is held fixed, the
phase is the ``electron" Aharonov-Bohm phase $\Phi/\Phi_0$, and if the
filling fraction is held fixed, the phase is the ``quasiparticle"
Aharonov-Bohm phase $\Phi/\Phi^*$.

In the absence of tunneling, the current $I$ equals the Hall current
$I_H=\nu e^2/h V$. In the presence of tunneling, the transmission
current $I$ satisifies $I=I_H-I_t$, where $I_t$ is the tunneling
current. Treating the tunneling term perturbatively in the model
above, we can solve for the tunneling current $I_t$ as a function of
the voltage $V$ between the edges, to low orders in the tunneling
amplitude. This perturbative result is valid as long as $I_t$ is small
compared to the Hall current $I_H$. It is easy to generalize the
problem to $N$ contacts located at $x_i$ with tunneling amplitudes
$\Gamma_i$, for $i=1,..,N$, and in appendix
\ref{appertshort} we solve this general problem. The result can be cast 
in a form very similar to the one point-contact result.  At zero
temperature, it is given by
\begin{equation}
I_t=e^* |\Gamma_{\rm eff}|^2 \ {2\pi \over \Gamma(2g)}
|\omega_J|^{2g-1}\ {\rm sign}(\omega_J)\ ,
\end{equation}
where, for several point-contacts, $\Gamma_{\rm eff}$ is the effective
coupling which includes the interference between the couplings
$\Gamma_i$, $i=1,..,N$. The effective coupling is given by
\begin{equation}
|\Gamma_{\rm eff}|^2=\sum_{i,j=1}^N \Gamma_i\ \Gamma^*_j\ H_g(\omega_J
|x_i-x_j|)\ ,
\end{equation}
with
\begin{equation}
H_g(x)=\sqrt{\pi}\ {\Gamma(2g) \over \Gamma(g)}\ 
{J_{g-1/2}(x) \over (2x)^{g-1/2}}\ ,
\end{equation}
where $J_{g-1/2}$ is a Bessel function of the first kind. In the case
of a single point-contact, the effective coupling is given by
$\Gamma_{\rm eff}=\Gamma$, independent of
frequency, and we recover the familiar results of Ref. \cite{XGW2}.

In the case of the two point-contact geometry, we have an effective
coupling
\begin{equation}
|\Gamma_{\rm eff}|^2=|\Gamma_1|^2 + |\Gamma_2|^2 + 
(\Gamma_1\Gamma^*_2+\Gamma^*_1\Gamma_2)\ H_g({\omega_Ja\over v})\ ,
\label{pert-Gammaeff}
\end{equation}
where $a=|x_1-x_2|$ is the linear distance along the edge between the
contacts, and we have restored the velocity $v$ to the equation. (If
the path length between the point-contacts for the left and right
edges are different, then $a$ is the average of the two lengths and
there is an extra contribution to the relative phase between
$\Gamma_1$ and $\Gamma_2$.)  The separation $a$ sets the time scale
$\tau=a/v$, and thus the frequency scale $\omega_{\rm osc}=2\pi/\tau$
for oscillations in the value of the effective coupling $\Gamma_{\rm
eff}$. It is easy to check that $H_g(x)\rightarrow 1$ as $x\rightarrow
0$, and that $H_g(x)\rightarrow 0$ as $x\rightarrow \infty$, so that
the effective coupling has the asymptotic values

\[ |\Gamma_{\rm eff}|^2=\left\{ \begin{array}{ll}
\ |\Gamma_1+\Gamma_2|^2 \ ,&\mbox{$\omega_J \ll \omega_{\rm osc}$}\\
\\
|\Gamma_1|^2 + |\Gamma_2|^2 ,&\mbox{$\omega_J \gg \omega_{\rm osc}$}
\end{array}
\right. \]
which correspond to coherent and incoherent interference between
$\Gamma_1$ and $\Gamma_2$. In the first case $\Gamma_{\rm eff}$
depends on the relative phase between $\Gamma_1$ and $\Gamma_2$, so
the tunneling current should clearly exhibit the Aharonov-Bohm
oscillations, and in the second case the Aharonov-Bohm effect is
washed out.

\begin{figure}
\vspace{1cm}
\hspace{1.3in}
\epsfxsize=4in
\epsfbox{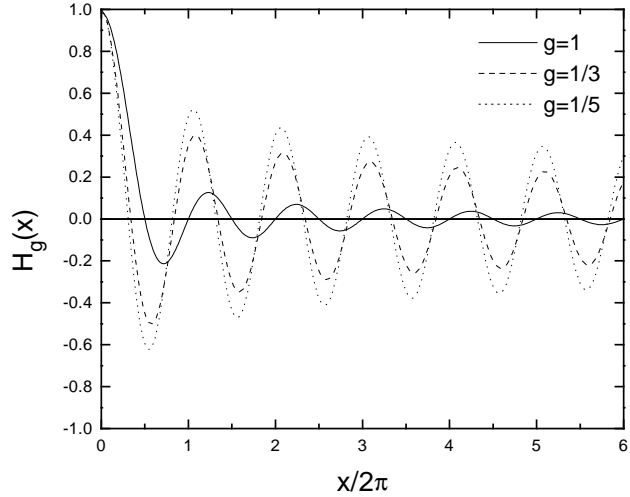}
\vspace{.5cm}
\caption{Modulation $H_g(x)$ for $g=1,1/3,1/5$. Notice that the
decay rate of the modulation is $x^{-g}$. Also, the position of the
zeros of $H_g(x)$ (those of $J_{g-1/2}(x)$) are approximately given by
$x_n\approx\pi(n+\xi_g)$, $n$ integer, where $\xi_g=\frac{1+g}{2}$ is
a $g$ dependent shift}
\label{fig2}
\end{figure}

For the intermediate range of frequencies comparable to $\omega_{\rm
osc}$, there will be oscillations in the effective coupling as a
function of $\omega_J$.  This interference term depends on the
relative phase between $\Gamma_1$ and $\Gamma_2$, which can be
adjusted by varying the magnetic flux $\Phi$ through the area between
the two point-contacts.  If an experimental aparatus is setup to
detect the component of the current that oscillates with the flux
$\Phi$, the magnitude of the oscillations will be
\begin{equation}
|I_t^\Phi|=e^*|\Gamma_1||\Gamma_2| {2\pi \over \Gamma(2g)}
\ |\omega_J|^{2g-1}\ |H_g({\omega_Ja\over v})|\ .
\end{equation}
The behavior of $|I_t^\Phi|$ as a function of the source-drain voltage
can be understood by looking at the plot of $H_g(x)$, for different
$g$, in Figure \ref{fig2}. The envelope of the decaying oscillations
in $H_g(x)$ is algebraic ($\sim x^{-g}$), so that the envelope of
$|I_t^\Phi|$ scales as $\omega_J^{g-1}$, or in other words,
$|I_t^\Phi|\propto V^{g-1}$.

The zeros of $H_g(x)$ are those of the Bessel function $J_{g-1/2}(x)$,
which are separated by a distance approximately equal to $\pi$. The
zeros are approximately given by
\begin{equation}
x_n\approx \pi(n+\xi_g)\ ,\ \ 	n=0,1,\dots \qquad {\rm with} 
\qquad
\xi_g=\frac{1+g}{2}\ ,
\end{equation}
which fits the graphs in Figure \ref{fig2} very well. Notice that for
$g=1$, the position of the zeros is exactly given by $x_n=\pi(n+1)$,
$n=0,1,\dots$, and all zeros are equally spaced. The ratio $x_1/x_0$
between the first two zeros is 2. For $g\ne1$, even though the zeros
are approximately equally separated, they are shifted, and
$x_1/x_0\approx \frac{g+3}{g+1}\ne 2$. This example illustrates how
the position of the zeros can be used to probe Luttinger liquid
behavior.

Experimentally, the position of these nodes can be observed precisely
by measuring those source-drain voltages for which the Aharonov-Bohm
oscillations simply disappear. We would like to stress that
the experimental measurement of the position of the nodes of the
interference patterns provides a clean (interferometric) way of
probing Luttinger liquid behavior. Such an interferometric measurement
focuses on the behavior between the two contacts, and may be free of
parasitic effects elsewhere that can mask the anomalous scaling
behavior of Luttinger liquids.

Finally, the location of the nodes could also provide another 
method of measuring fractional charge.  The position of the nodes 
depends on $\omega_Ja/v$, where the Josephson frequency $\omega_J = 
e^* V/\hbar$ depends on the fractional charge.  If there were an 
independent measurement of the velocity the of edge modes $v$, then 
the location of the nodes would yield a value for $e^*$.


\section{Free Fermions}
\label{secg=1}

To better understand the behavior of our perturbative solution for the
current, we will look at the exact solution for the $g=1$ case with
two constrictions. This case reduces to a simple problem that is 
essentially the same as 
wave transmission through a Fabry-Perot interferometer. We
can solve for the transmission and reflection amplitudes at a
single impurity, which, for $g=1$, are independent of the energy of
the incident waves. Then, in addition to using these amplitudes to
account for the scattering at  each of the two impurities,
we must also propagate the waves from one impurity
to another, which is the part responsible for the interference
effects.  Each frequency $\omega$ will then be transmitted with a 
different
coefficient $T(\omega)$. For an applied voltage $V$ between the
edges, there will be a whole range of frequencies of width
$\omega_J=eV/\hbar$ in the incident wave packet.  We must then
integrate the final transmission coefficients over the energies in
this band that contribute to the total current.

Consider, to begin with, a single point-contact with tunneling
amplitude $\Gamma$, which can either transmit or scatter a particle.
For $g=1$, we can work in terms of free fermions given by $\psi_{R,L}={1
\over \sqrt{2\pi}} e^{\pm i\phi_{R,L}(x,t)}$, with Hamiltonian
\begin{eqnarray}
H = \int d&x& \bigg\{{\psi_R}^\dagger(x)\left[-i{\partial \over\partial x}-
       {\omega_0\over2}\right]\psi_R(x) +
        {\psi_L}^\dagger(x)\left[i{\partial \over\partial x}+
       {\omega_0\over2}\right]\psi_L(x) \nonumber \\
   &+& 2\pi\delta(x)\left[\Gamma_i {\psi_L}^\dagger(x) \psi_R(x) + 
H.c.\right]
       \bigg\}. \label{freeH}
\end{eqnarray}
Once again, we have set the velocity $v$ of the chiral fermions to 1.
It is then a straightforward wave mechanics problem to solve for the
scattering matrix, $S$, which gives the transformation from the
incoming modes, $\tilde\psi_{R-}(\omega)$, $\tilde\psi_{L+}(\omega)$
to the outgoing modes, $\tilde\psi_{R+}(\omega)$,
$\tilde\psi_{L-}(\omega)$.  We find that $S = \pmatrix{t_i & r_i\cr
-{r_i}^* & t_i \cr}$, where the transmission and reflection amplitudes
are given by
\begin{equation}
t_i= {1 - \pi^2 |\Gamma_i|^2 \over 1 + \pi^2 |\Gamma_i|^2}
\qquad {\rm and} \qquad
r_i=  {-i 2\pi {\Gamma_i}^* \over 1 + \pi^2 |\Gamma_i|^2}\ .\label{t-r}
\end{equation}
The transmission and reflection coefficients are $T=|t|^2$ and
$R=|r|^2$, respectively. (We note here that $\Gamma_i$ in Equation
\ref{t-r} is renormalized so that the strong barrier limit or
total reflection ($R=1$) occurs when $|\Gamma_i| = 1/\pi$.  This is in
contrast with the case when $g=1/2$ or $1/3$, when total reflection
occurs as $\Gamma \to \infty$. Technically, this difference arises
because for $g<1$ the tunneling operator is infrared relevant whence
short distance behavior does not matter, whereas for $g=1$ it is marginal.)

When there is more than one scatterer, it is convenient to use the
transmission matrix approach to find the transmission through and
reflection out of the two point-contacts separated by a distance $a$.
Recall that the transmission matrix $M_i$ gives the transformation
from the states on the left of the barrier, $\tilde\psi_{R-}(\omega)$,
$\tilde\psi_{L-}(\omega)$, to the states on the right of the barrier,
$\tilde\psi_{R+}(\omega)$, $\tilde\psi_{L+}(\omega)$, and can be
obtained directly from $S$.  After passing through the first
constriction, the waves propagate to the right a distance $a$, which
results in multiplying the modes $\pmatrix{\tilde\psi_R(\omega) \cr
\tilde\psi_L(\omega)\cr}$ by $D = \pmatrix{e^{ia\omega} & 0 \cr 0 &
e^{-ia\omega}\cr }$.  Finally, the waves are scattered again by the
second point-contact, so the waves on the right-hand side depend on
the waves to the left of the scatterers as follows:
\begin{equation}
\pmatrix{\tilde\psi_{R+}(\omega) \cr \tilde\psi_{L+}(\omega)\cr}
    = M_2 D M_1 
    \pmatrix{\tilde\psi_{R-}(\omega) \cr \tilde\psi_{L-}(\omega)\cr}.
  \label{MDM}
\end{equation}
Note that the matrix $D$ contains the phase $e^{\pm i\omega x}$, and 
it is this phase which is responsible for the voltage oscillations.  
In particular, the right-moving and left-moving modes that scatter between
the two point-contacts have opposite phases which interfere with each other.
Multiplying out the matrices in Eq. (\ref{MDM}), we find that the
transmission amplitude for the two point-contact geometry is
\begin{equation}
t(\omega)=\frac{t_1t_2}{1+r_1r^*_2e^{2i\omega a}}\ ,
\end{equation}
where $t_{1,2}$ and $r_{1,2}$ are the transmission and reflection
amplitudes for the two contacts, as given by Eq. (\ref{t-r}).
The transmission coefficient $T(\omega)$ through the whole droplet is then
\begin{equation}
T(\omega)=\frac{|t_1|^2|t_2|^2}
{1+|r_1|^2|r_2|^2+(r_1r^*_2e^{2i\omega a}+r^*_1r_2e^{-2i\omega a})}\ .
\label{T-omega}
\end{equation}

With this frequency dependent transmission coefficient, we can
calculate the current for an energy difference $\omega_J$ between the
right and left moving edges. It is given by the total right-moving
current minus the total left-moving current passing through a point
$x$.  If we choose the point to be to the right of the barrier, then
the total right-moving current at energy $\omega$ is given by the
transmitted right-moving current $e{\cal T}(\omega)n^R(\omega)$ plus
the total left-moving current that was reflected into right movers
$e[1 - {\cal T}(\omega)]n^L(\omega)$. Similarly, the total left-moving
current at energy $\omega$ is just the total incoming left-moving
current $n^L(\omega)$, where $n^{R,L}(\omega)$ are the occupation
numbers of right and left movers. In this simple model we can easily
include the temperature dependence of the transmission because it can
be completely accounted for by the Fermi-Dirac distribution of the
left and right movers:
\begin{equation}
n^{R,L}(\omega)=\frac{1}{e^{\beta(\omega\mp\omega_J/2)}+1}\ .
\end{equation}
The expression for the current through the droplet then reduces to
\begin{equation}
I = e \int_{-\infty}^{\infty}{d\omega \over 2\pi}
T(\omega)\left[n^R(\omega) - n^L(\omega)\right]\ .\label{g=1-int}
\end{equation}
At zero temperature, the integral in Eq. (\ref{g=1-int}) yields
\begin{equation}
I={e\over 2\pi a}{|t_1|^2|t_2|^2 \over 1-|r_1|^2|r_2|^2}
\tan^{-1}\left(
\frac{{1-|r_1|^2|r_2|^2 \over 1+|r_1|^2|r_2|^2}\sin(\omega_Ja)}
{\cos(\omega_Ja) + {r_1r^*_2+r^*_1r_2 \over 1+|r_1|^2|r_2|^2}}
\right)\ .  \label{g=1T=0-It}
\end{equation}
Notice that if $t_1=t_2=1$ (total transmission), $I={e\over
2\pi}\omega_J={e^2\over h}V$. For small tunneling amplitutes
$\Gamma_1$ and $\Gamma_2$ ($|r_1|^2,|r_2|^2\ll 1$), we find that
$I={e^2\over h}V-I_t$, where
\begin{equation}
I_t=e^* |\Gamma_{\rm eff}|^2 \ 2\pi \omega_J\ ,
\label{I_t-g=1}
\end{equation}
with
\begin{equation}
|\Gamma_{\rm eff}|^2=|\Gamma_1|^2 + |\Gamma_2|^2 + 
(\Gamma_1\Gamma^*_2+\Gamma^*_1\Gamma_2)
\ {\sin(\omega_J a)
\over (\omega_J a)}\ .
\end{equation}
This is the same as the result obtained perturbatively in section
\ref{secModel} if we set $g=1$ in Eq. (\ref{pert-Gammaeff}).

If we expand the transmission coefficient $T(\omega)$ in
Eq. (\ref{T-omega}) for small tunneling amplitudes, 
we can easily obtain the finite temperature
tunneling current $I_t$. It is still given by Eq. (\ref{I_t-g=1}),
but the effective coupling is now
\begin{equation}
|\Gamma_{\rm eff}|^2=|\Gamma_1|^2 + |\Gamma_2|^2 + 
(\Gamma_1\Gamma^*_2+\Gamma^*_1\Gamma_2)
\ {2\pi T a\over \sinh(2\pi T a)}{\sin(\omega_J a)
\over (\omega_J a)}\ .
\end{equation}
Notice that the distance $a$ sets the temperature scale for which the
interference term $\Gamma_1\Gamma^*_2+\Gamma^*_1\Gamma_2$ decays.

In the general case, such as for other filling fractions, the current
should still be obtainable by an expression like Equation \ref{g=1-int},
where $T(\omega)$ is the transmission coefficient and $n^R(\omega)$ and
$n^L(\omega)$ are the number densities of filled states at energy $\omega$.
If the behavior of the system deviates from the result in Equation 
\ref{g=1T=0-It}, this should indicate that the transmission and reflection
amplitudes for a single point-contact depend on energy and that the
density of states no longer has the simple Fermi-liquid form.


\section{Finite Temperature Effects}
\label{secfiniteT}

In section \ref{secModel} we have found that, at zero temperature, the
effect of the two point-contacts can be completely absorbed into an
effective coupling $\Gamma_{\rm eff}$, which describes all the
interference between the two contacts. In this section we will show
that when $T\ne 0$, this is still the case, but now $\Gamma_{\rm eff}$
will depend on temperature also.

The finite temperature $T$ brings another energy scale to the problem.
This energy scale should be compared to the one set by the separation
between the contacts $a$, which is given by $\omega_{\rm osc}=2\pi
v/a$. Thus, when $k T \gg \hbar\omega_{\rm osc}$, the interference
effects should be washed out. One should also keep in mind the energy
scale associated with the Josephson frequency $\omega_J=e^*V/\hbar$,
so that the decay of the interference effects with temperature will
depend on the ratios of the three energy scales $T$, $\omega_J$ and
$\omega_{\rm osc}$. The interesting question to ask is how the
different $g$ affect the way the interference is washed out, or,
equivalently, how the filling factor $\nu$ of the underlying FQH state
affects the decay of the oscillations with temperature.

To lowest order in the tunneling amplitude $\Gamma$,
the tunneling current between edge states in the presence of a single
point-contact at finite temperature is given by \cite{XGW2}:
\begin{equation}
I_t=e^*\ |\Gamma|^2 
4(\pi T)^{2g-1}\ 
B\left(g-i{\omega_J \over 2\pi T},g+i{\omega_J \over 2\pi T}\right)
\sinh \left({\omega_J \over 2 T}\right)\ ,
\end{equation}
where $B$ is the beta function. In appendix \ref{appertshort} we show that
the same expression gives the current for $N$ point-contacts with
couplings $\Gamma_i$, $i=1,...,N$ if we use an effective coupling
\begin{equation}
|\Gamma_{\rm eff}|^2=\sum_{i,j=1}^N \Gamma_i\ \Gamma^*_j\ H_g(\omega_J,
|x_i-x_j|,T)\ ,
\end{equation}
with 
\begin{equation}
H_g(\omega_J,x,T)=2\pi{\Gamma(2g) \over \Gamma(g)} {e^{-2g\pi T|x|}
\over
\sinh {\omega_J \over 2 T}}\
\times\ {\rm Im}
\bigg\{
{e^{-i\omega_J |x|}\ F(g,g+i{\omega_J \over 2\pi T};
1+i{\omega_J \over 2\pi T};e^{-4\pi T|x|})
\over
\Gamma(g-i{\omega_J \over 2\pi T})\ \Gamma(1+i{\omega_J \over 2\pi T})}
\bigg\}\ .
\end{equation}
In this expression, $F$ is a hypergeometric function.
Notice that the function $H_g$ depends on $T$, $x$ and $\omega_J$ only
through the combinations $\omega_J x$ and $\omega_J/(2\pi T)$. We can
thus cast the modulation $H_g\left(\omega_J x,{\omega_J\over 2\pi
T}\right)$ in terms of the following function of two variables:
\begin{equation}
H_g(y_1,y_2)=2\pi{\Gamma(2g) \over \Gamma(g)} {e^{-g y_1/y_2}
\over
\sinh (\pi y_2)}\
\times\ {\rm Im}
\bigg\{
{e^{-i y_1}\ F(g,g+iy_2;
1+iy_2;e^{-2y_1/y_2})
\over
\Gamma(g-iy_2)\ \Gamma(1+iy_2)}
\bigg\}\ .\label{H_g(y_1,y_2)}
\end{equation}   
The effective coupling for a two point-contact geometry is then
\begin{equation}
|\Gamma_{\rm eff}|^2=|\Gamma_1|^2 + |\Gamma_2|^2 +
(\Gamma_1\Gamma^*_2+\Gamma^*_1\Gamma_2)
\ H_g\left({2\pi\omega_J \over \omega_{\rm osc}},{\omega_J\over 2\pi T}\right)\ .
\end{equation}
In this form, it is clear that the interference term depends on the
ratios of the three energy scales in the problem.

\begin{figure}
\vspace{1cm}
\hspace{1.3in}
\epsfxsize=4in
\epsfbox{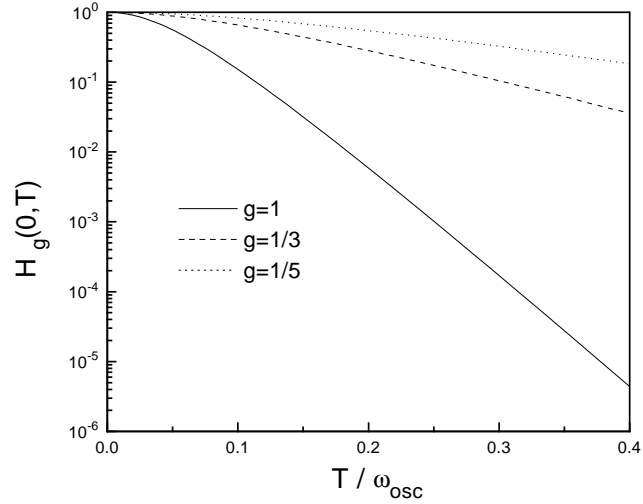}
\vspace{.5cm}
\caption{Temperature decay of the modulation $H_g$ for $g=1,1/3,1/5$.
The quantity plotted is $H_g(\omega_J=0,a,T)$ vs.  $T$, with $T$
measured in units of $\omega_{\rm osc}=2\pi \frac{v}{a}$ (the energy
scale associated with the point-contact separation $a$). Notice that
the modulation decays exponentially with $T$ for large temperatures,
and that the decay rate is faster for larger $g$.}
\label{fig3}
\end{figure}

We begin to explore how different values of $g$ change the behavior of
the modulation $H_g$ by considering Fermi liquid ($g=1$) edge
states associated with a QH filling factor $\nu=1$.  In this case,
Eq. (\ref{H_g(y_1,y_2)}) can be shown to simplify to
\begin{equation}
H_1(y_1,y_2)={y_1/y_2 \over \sinh(y_1/y_2)}
             {\sin y_1 \over y_1}\ ,
\end{equation} 
so that
\begin{equation}
|\Gamma_{\rm eff}|^2=|\Gamma_1|^2 + |\Gamma_2|^2 +
(\Gamma_1\Gamma^*_2+\Gamma^*_1\Gamma_2)
\ {4\pi^2 T / \omega_{\rm osc}
\over \sinh(4\pi^2 T /\omega_{\rm osc})}
\ {\sin(2\pi\omega_J / \omega_{\rm osc})
\over (2\pi\omega_J / \omega_{\rm osc})} \ .
\end{equation}
This is the same as the expression obtained in section \ref{secg=1}
directly from the free fermion transmission approach. Notice that for
$g=1$ the finite temperature correction appears only as a
multiplicative factor in front of the modulation for $T=0$. This is
not necessarily the case for other $g$, as shown below in Fig.
\ref{fig4}.  This multiplicative factor decays exponentially
($1/\sinh(4\pi^2 T /\omega_{\rm osc})$) with temperature, with the
scale ($\omega_{\rm osc}$) set by the two point-contact separation
$a$.  In Fig. \ref{fig3} we show the decay of $H_g$ with temperature
for $\omega_J=0$ in a log-plot. From this plot we can extract how the
modulation decays with temperature $T$ for different $g$.  Using
asymptotic expressions for the hypergeometric function, we find that
for $T \gg \omega_{\rm osc}$, the function
$H_g({2\pi\omega_J \over\omega_{\rm osc}},{\omega_J\over 2\pi T})$ decays as
$e^{-4\pi^2 g T/\omega_{\rm osc}}$, whereas for $T \ll
\omega_{\rm osc}$, the fall off is much slower.

Another interesting quantity is presented in Fig. \ref{fig4}, where we
display the ratio $H_g(\omega_J,a,T)/H_g(0,a,T)$ between the
modulation at $\omega_J$ and at $\omega_J=0$, for different
temperatures. The natural variables for displaying this dependence are
the ratios $T/\omega_{\rm osc}$ and $\omega_J/\omega_{\rm osc}$ (put
differently, we measure energies as compared to the scale $\omega_{\rm
osc}$ set by the separation $a$ between the contacts). Notice that,
for general $g$, the curves move around as a function of $T$. The
curves collapse into one only for $g=1$. Also notice that as the
temperature increases, the position of the zeros for $g=1/3$ and
$g=1/5$ approaches those for $g=1$, so that increasing temperature
masks the Luttinger liquid behavior, with a crossover temperature
roughly equal to $\omega_{\rm osc}$.

\begin{figure}
\hspace{1.8 in}
\epsfxsize=3in
\epsfbox{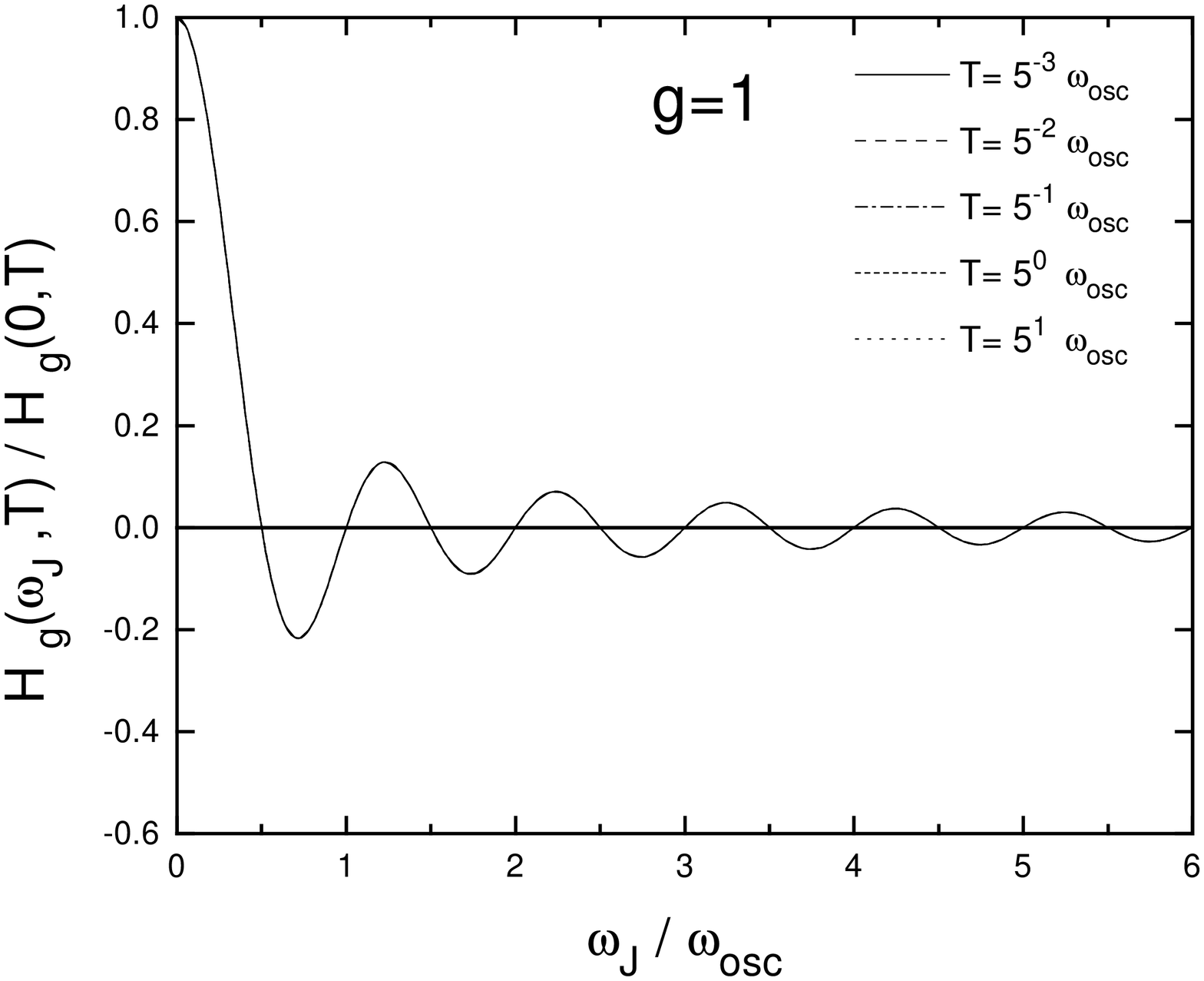}
\vspace{0.2in}
\end{figure}
\begin{figure}
\hspace{.17 in}
\epsfxsize=3in
\epsfbox{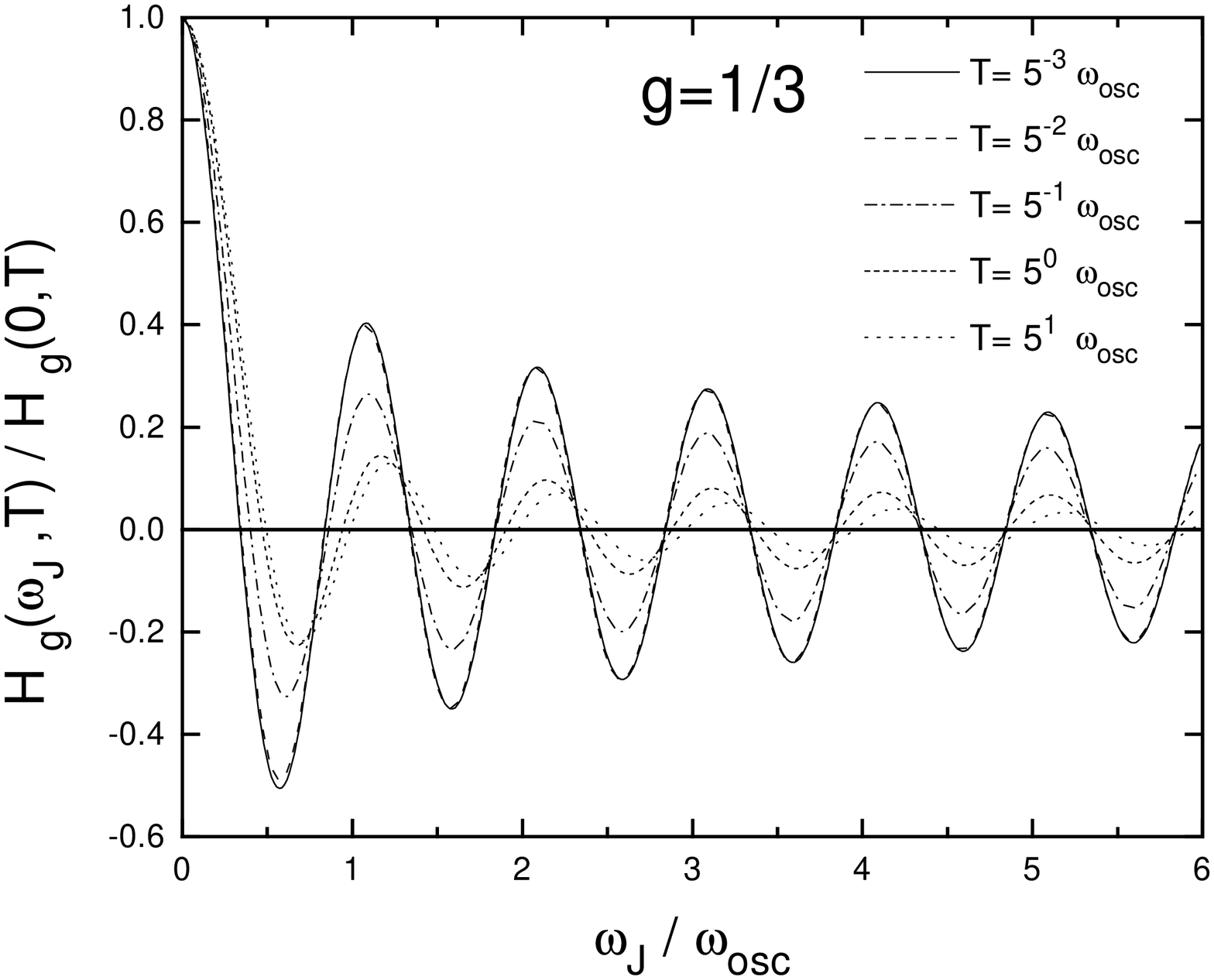}
\hspace{.15 in}
\epsfxsize=3in
\epsfbox{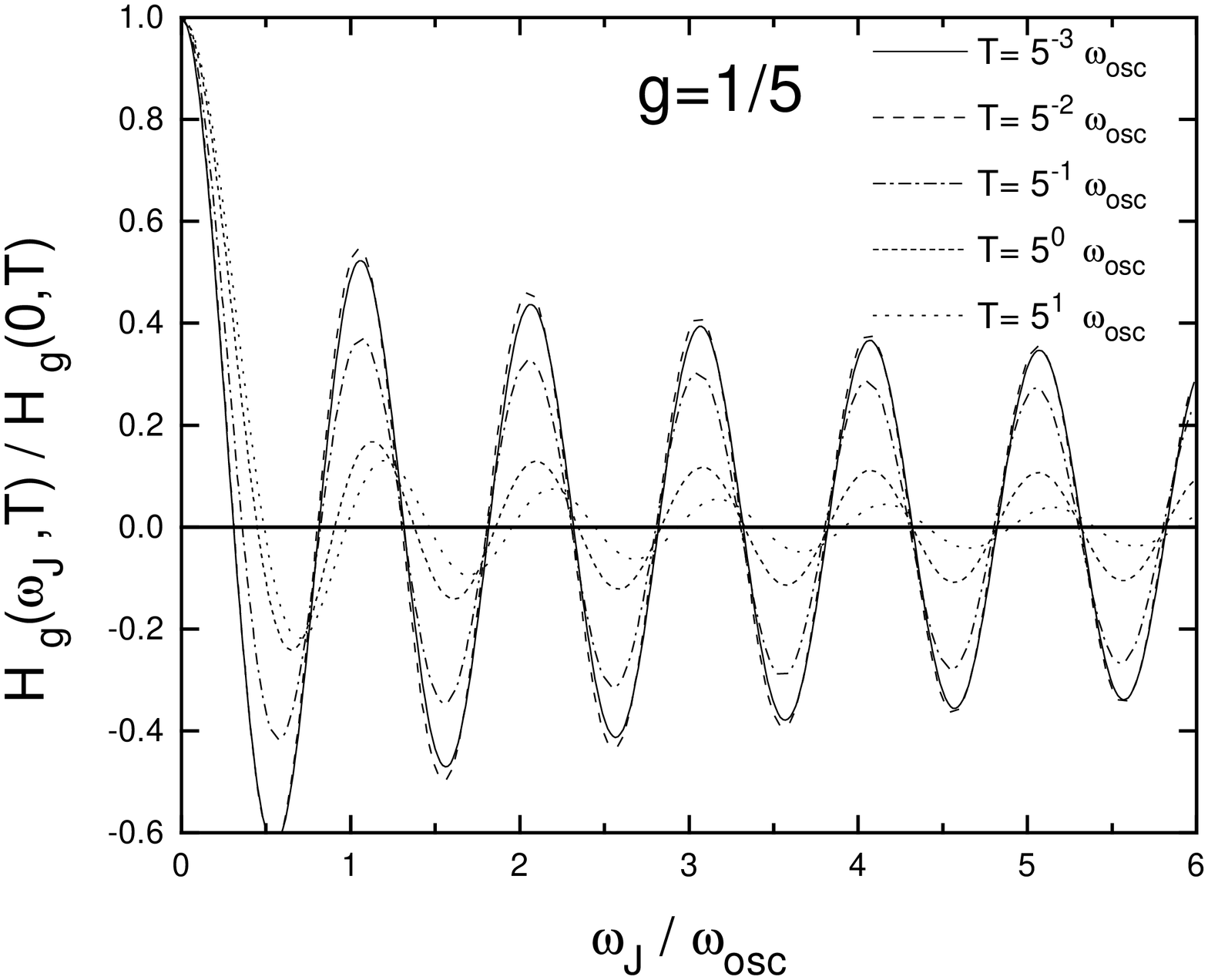}
\vspace{1cm}
\caption{Dependence of the modulation $H_g(\omega_J,a,T)$ on
$\omega_J$ for different $T$. The quantity plotted is the rescaled
$H_g$ ($H_g(\omega_J,a,T)/H_g(0,a,T)$) so as to show how the shape of
the modulation curve changes with $T$ for different $g$. Notice that
all curves collapse for $g=1$, {\it i.e.}, all frequencies get
suppressed uniformly as temperature is increased. For $g=1/3$ and
$g=1/5$, however, notice that the curves do not collapse together
anymore, and that higher $\omega_J$ are supressed more strongly than
lower $\omega_J$ as $T$ is increased.}
\label{fig4}
\end{figure}


\section{Numerical Estimates}
\label{estimates}

In this section, we will give estimates of the sizes of parameters at which
the interference effects could be observed.  First, we will consider the
change in magnetic field, $\Delta B$, required for one Aharonov-Bohm period,
which is given by 
\begin{equation}
\Delta B = {\ e\ \over \ e^*} {\Phi_0\over A} 
= {\ e\ \over \ e^*} {41\mu {\rm m}^2 
\ {\rm gauss} \over A}.
\end{equation}
If the number of electrons is held constant, then in this equation
$\Phi_0$ is equal to one flux quantum and $e^*$ is the charge of an
electron, $e$.  If, instead, the filling fraction is held fixed, then
$\Phi_0$ is $e/e^*$ times one flux quantum, where $e^*$ is the charge
of the quasiparticle.  In this equation, $A$ is the area of the FQH
liquid between the two contacts, and is roughly given by $A = ad$,
where $d$ is the width of the sample and $a$ is the distance between
the two contacts.  If we assume the width is $d = 1 \mu {\rm m}$, then
the period is related to the distance between the contacts by
\begin{equation}
\Delta B = \cases{41  \mu{\rm m-gauss}/a 
                   & for fixed number of electrons $n$ \cr
                120 \mu{\rm m-gauss}/a
                   & for fixed filling fraction $\nu= 1/3$.}
\end{equation} 
For $a = 1 \mu {\rm m}$ or $10 \mu {\rm m}$, the ``electron"
Aharonov-Bohm period is $\Delta B = 41$ gauss and $4.1$ gauss,
respectively, and the ``quasiparticle" Aharonov-Bohm period is $\Delta
B = 120$ gauss and $12$ gauss, respectively.

Next, we consider the voltage fluctuations.  The separation between the
zeros of the Bessel function $J_{g-1/2} (x)$ is approximately equal to 
$\pi$
and the location of the nodes in the voltage fluctuations roughly occur
when
\begin{equation}
{\omega_J a \over v} \approx \pi(n+\xi_g)\ ,\ \ n=0,1,\dots \qquad
{\rm with} \qquad
\xi_g=\frac{1+g}{2}\ ,
\end{equation}
As noted earlier, depending on the value of $g$, the precise location of
these nodes will be shifted a little, which may provide a way of 
distinguishing between Luttinger liquid behavior and other types of
behavior. In this equation, we have restored the velocity of the edge
modes $v$, which earlier was set to 1.  For $g = 1/3$, an estimate for 
$v$ \cite{Moon} is $v \approx 10^5 {\rm m}/{\rm s}$.  Using 
$\omega_J = e^* V/\hbar$, we find that the voltage at the nodes and the
distance between the point-contacts must satisfy
\begin{equation}
Va = (n +\xi_g)  {e\over e^*} \times 200 \mu{\rm V}-\mu{\rm m}, 
\end{equation}
where $V$ has units of microvolts and $a$ has units of microns. For
$\nu = 1/3$, we take $e^* / e = 1/3$.  Thus, for $a = 1 \mu {\rm m}$
or $10 \mu{\rm m}$ the voltage at the first node is roughly $400
\mu{\rm V}$ and $40 \mu{\rm V}$, respectively. Lastly, we will
estimate the coherence length, or the amount by which the temperature
reduces the signal.  However, we note that phonons can also lead to
dephasing, althought we do not consider them here. For temperatures
greater than $kT > \hbar\omega_{\rm osc}/(4\pi^2)$, the interference
effects fall off as $Te^{-4\pi^2 g kT/(\hbar\omega_{\rm osc})}$, where
$\omega_{\rm osc} = 2\pi v/a$.  Thus, for
\begin{equation}
kT < {\hbar v\over 2\pi g a}
\end{equation}
the signal is not affected much by temperature.  We can define the
coherence length, $a_c$ by the spacing for which the signal has
decreased roughly by a factor of $1/e$, so that $a_c = v\hbar/(g 2\pi
kT)$. Then, for $v \approx 10^5 {\rm m}/{\rm s}$ and $ g = 1/3$, at $T
= 100 {\rm mK}$ the coherence length is $a_c = 4 \mu{\rm m}$, and for
$T = 30 {\rm mK}$, the coherence length is $a_c = 12 \mu {\rm m}$.
Thus, the signal for a separation of $1 \mu {\rm m}$ should not be
noticeably affected at either temperature, and even for a separation
of $10\mu {\rm m}$ the signal will be attenuated by a factor of 2 at
$30 {\rm mK}$ and by a factor of 15 at $100 {\rm mK}$.

It follows then that a separation of a few microns should be
sufficient to allow observation of the inteference effects at
temperatures around and below $100 {\rm mK}$. The requirement on the
gates is that they be close enough that their electrostatic
``shadows'' do not overlap in the plane of the electron gas. For gate
diameters of about $1000 \AA$, this condition could be met by placing
them about $1000 \AA$ from the electron gas.


\section{Conclusions}
\label{conclusion}

In this paper we have proposed a device, the two point-contact 
interferometer, consisting of a Hall bar with two weak barriers, 
that can be used to study quantum interference effects in a strongly
correlated system. The device allows for the study of three types of 
interference effects: Aharanov-Bohm oscillations with magnetic field,
statistical oscillations with quasiparticle number and
Fabry-Perot oscillations with source-drain voltage. These interference 
effects can be used to measure the fractional charge and statistics of
quasiparticles in the quantum Hall effect. They also provide a new way
of searching for non-Fermi liquid behavior in the dynamics of the edges.
We would like to emphasize that much of our account of the physics 
of the device is quite robust, in that it depends upon quite general 
``topological'' properties of QH quasiparticles; our proposals for 
measuring charge and statistics fall in this category. Other features,
such as the details of the Fabry-Perot nodes are more specific to the
simplest version of edge state dynamics used in the calculations and as 
such are subject to the caveat that they
represent the behavior of the system only at the lowest energies.


\begin{center}
{\bf ACKNOWLEDGEMENTS}
\end{center}

We would like to thank several colleages who gave us important
suggestions and constructive criticism on the experimental
implications of this work: David Abusch-Magder, Ray Ashoori, Marc
Kastner, Beth Parks and Nikolai Zhitenev at MIT; Hari Manoharan,
Kathryn Moler, Dan Shahar, Mansour Shayegan and Lydia Sohn at
Princeton; Hong-Wen Jiang at UCLA. SLS would like to thank Phillip
Phillips for introducing him to resonances in interacting systems and
for suggesting that the two barrier problem might be a good thing to
look at. This work is supported by NSF grants DMR-94-00334 (CCC),
DMR-93-12606 (SAK) and DMR-94-11574 (XGW). XGW and SLS acknowledge
support from the A. P. Sloan Foundation.  D.~F.~is currently a Bunting
Fellow sponsored by the Office of Naval Research.


\appendix
\section{perturbative calculation}
\label{appertshort}

We derive here the correction to the Hall current due to tunneling at
the point-contacts. We will assume the general case of $N$ contacts at
locations $x_i$ and coupling $\Gamma_i$, $i=1,...,N$.

The first step in the calculation is to obtain the tunneling current
operator $j(t)$. This operator includes the tunneling currents flowing
from one edge to the other through all $N$ point-contacts in the
problem. The tunneling operator can be obtained from the time
evolution of the total charge operators $Q_{R,L}$ on the $R,L$ edges:
\begin{equation}
j(t)=-\frac{1}{i\hbar}[Q_L,H]=\frac{1}{i\hbar}[Q_R,H]\ .
\end{equation}
The charge operator commutes with the free part of the Hamiltonian, so
that the only contribution comes from the tunneling term
\begin{equation}
H_{\rm tun}=\sum_{i=1}^N \Gamma_i
e^{-i\omega_J t}e^{i\sqrt{g}\phi(t,x_i)}+ H.c.\ .
\end{equation}
Using the commutation relations
for the bosonic fields $\phi_{R,L}$, we obtain
\begin{equation}
j(t)=ie^* \sum_{i=1}^N \Gamma_i e^{i\sqrt{g}\phi(t,x_i)}+H.c. \ .
\end{equation}

The expectation value for the current at time $t$ is given by
\begin{equation}
\langle j(t)\rangle =
\langle 0|S^\dagger(t,-\infty)\ j(t)\ S(t,-\infty) |0\rangle\ ,
\end{equation}
where $S(t,-\infty)$ is the time evolution operator. The next step is
to calculate $\langle j \rangle$ perturbatively in the tunneling
amplitudes $\Gamma_i$. Because there is a voltage difference between
the $R$ and $L$ terminals, the system is out of thermodynamical
equilibrium, and we must to use field theoretical tools appropriate
for such non-equilibrium problems \cite{CFW}. However, non-equilibrium
effects appear only to second and higher orders in perturbation
theory. Because we will calculate the tunneling current only to first
order in perturbation theory, we will not have to use non-equilibrium
field theory in this particular calculation.

To lowest order in the tunneling perturbation we have
\begin{equation}
\langle j(t)\rangle =-i\int_{-\infty}^t\ dt'\
\langle 0|[j(t),H_{\rm tun}(t')]|0\rangle \ .
\label{pert1}
\end{equation}
In the calculation of
\begin{eqnarray}
\langle 0|j(t)\  H_{\rm tun}(t')|0\rangle &=&e^*\sum_{i=1}^N \sum_{j=1}^N
\ \langle 0|(i\Gamma_i e^{-i\omega_J t} e^{i\sqrt{g}\phi(t,x_i)}-
i\Gamma_i^* e^{i\omega_J t} e^{-i\sqrt{g}\phi(t,x_i)})\nonumber\\
&\ &\ \ \ \ \ \ \ \ \ \ \times\ (\Gamma_j e^{-i\omega_J
t'}e^{i\sqrt{g}\phi(t',x_j)} +\Gamma_j^* e^{i\omega_J
t'}e^{-i\sqrt{g}\phi(t',x_j)})|0\rangle
\end{eqnarray}
the non-vanishing terms are those that transfer zero total charge when
applied to $|0\rangle$. We then have
\begin{eqnarray}
&\ &\langle 0|j(t)\  H_{\rm tun}(t')|0\rangle = \nonumber \\
&\ & = ie^* \sum_{i,j=1}^N \Big(\Gamma_i\Gamma_j^*\ e^{-i\omega_J (t-t')}
\langle 0|e^{i\sqrt{g}\phi(t,x_i)}
e^{-i\sqrt{g}\phi(t',x_j)}|0\rangle - \Gamma_i^*\Gamma_j\  
e^{i\omega_J (t-t')} \langle
0|e^{-i\sqrt{g}\phi(t,x_i)} e^{i\sqrt{g}\phi(t',x_j)}|0\rangle\Big)
\nonumber\\ 
&\ & = ie^*\sum_{i,j=1}^N \ (\Gamma_i\Gamma_j^*\ e^{-i\omega_J (t-t')} 
- \Gamma_i^*\Gamma_j\ e^{i\omega_J (t-t')})
\ e^{g\langle 0|\phi(t,x_i)\phi(t',x_j)|0\rangle}\ .
\end{eqnarray}
The $\phi$ field correlation is 
\begin{eqnarray}
\langle 0|\phi(t,x)\phi(0,0)|0\rangle&=&
\langle 0|\phi_R(t,x)\phi_R(0,0)|0\rangle + 
\langle 0|\phi_L(t,x)\phi_L(0,0)|0\rangle\\
&=& -\ln [\delta +i(t-x)]-\ln [\delta +i(t+x)]\ ,\nonumber
\end{eqnarray}
where $\delta$ is an
ultraviolet cut-off scale. Let us define
\begin{equation}
P_g(t,x)=e^{g\langle 0|\phi(t,x)\phi(0,0)|0\rangle}
=\big[\delta + i(t+x)\big]^{-g}\times
\big[\delta + i(t-x)\big]^{-g}\ .
\end{equation}
Notice that $P_g(t,x)=P_g(t,-x)$. Using the expression above, we can write
\begin{eqnarray}
-i\langle [j(t),H_{\rm tun}(t')]\rangle &=&e^*\sum_{i,j=1}^N\ 
\left(\Gamma_i\Gamma_j^*\ e^{-i\omega_J (t-t')} 
- \Gamma_i^*\Gamma_j\ e^{i\omega_J (t-t')}\right) \nonumber\\ 
&\ &\ \ \times \ \Big(
P_g(t-t',x_i-x_j)-P_g(-t+t',x_i-x_j)
\Big)\ .
\end{eqnarray}
Inserting the above expression into Eq.(\ref{pert1}) and performing the
$t'$ integration, we obtain the current expectation value:
\begin{eqnarray}
\langle j(t)\rangle &=&e^* \sum_{i,j=1}^N\ 
{\Gamma_i\Gamma_j^*+\Gamma_i^*\Gamma_j \over 2}\ 
[{\tilde P_g}(\omega_J,x_i-x_j)-{\tilde P_g}(-\omega_J,x_i-x_j)]\ ,
\end{eqnarray}
where ${\tilde P}_g(\omega_J,x)$ is the Fourier transform of the
$P_g(t,x)$ with respect to time. The problem is then reduced to the
calculation of the ${\tilde P}_g$'s. It is easy to calculate
\begin{equation}
{\tilde P_g(\omega,0)}=
\int_{-\infty}^\infty dp \frac{e^{i\omega p}}{(\delta +ip)^{2g}}
=\frac{2\pi}{\Gamma(2g)}|\omega|^{2g-1}e^{-|\omega|\delta} \ \theta
(\omega) \ ,
\end{equation}
and we can express the case $x\ne 0$ in terms of the ${\tilde
P_g(\omega,0)}$:
\begin{eqnarray}
{\tilde P_g(\omega,x)}&=&
\int_{-\infty}^\infty {d\omega' \over 2\pi}\ 
P_{g/2}(\omega',0)\ P_{g/2}(\omega-\omega',0)
\ e^{-i(2\omega'-\omega)x}\\
&=&\theta(\omega)\ 
\int_{0}^{|\omega|} {d\omega' \over 2\pi}\ 
\omega'^{g-1}\ (\omega-\omega')^{g-1}
\ e^{-i(2\omega'-\omega)x}\nonumber\\
&=&{\tilde P}_{g}(\omega,0)\ H_g(\omega x)\ ,\nonumber
\end{eqnarray}
where
\begin{equation}
H_g(y)=\sqrt{\pi}\ {\Gamma(2g) \over \Gamma(g)}\ 
{J_{g-1/2}(y) \over (2y)^{g-1/2}}\ .
\end{equation}
The tunneling current between the edge states $I_t=\langle j(t) \rangle$
is then simply
\begin{equation}
I_t=e^*{2\pi \over \Gamma(2g)}
|\omega_J|^{2g-1}\ {\rm sign}(\omega_J)\ 
\sum_{i,j=1}^N \Gamma_i\Gamma_j^*\ H_g(\omega |x_i-x_j|)\ .
\end{equation}
The expression for the tunneling current can be cast exactly in the
same form as that for a single contact,
\begin{equation}
I_t=e^* |\Gamma_{\rm eff}|^2 \ {2\pi \over \Gamma(2g)}
|\omega_J|^{2g-1}\ {\rm sign}(\omega_J)\ ,
\end{equation}
but with an effective coupling $\Gamma_{\rm eff}$ due to the
interference between $\Gamma_i$, $i=1,..,N$ of the $N$ contacts:
\begin{equation}
|\Gamma_{\rm eff}|^2=\sum_{i,j=1}^N \Gamma_i\ \Gamma^*_j\ H_g(\omega_J
|x_i-x_j|)\ .
\end{equation}

The calculations for $T=0$ can be extended for finite temperature.
Basically, the algebraic correlations at $T=0$ are mapped to the
correlations at $T\ne 0$ by a conformal transformation \cite{Shankar}:
\begin{equation}
\frac{1}{[\delta + i(t\pm x)]^g}
\rightarrow
\left[\frac{\pi T}{\sin\left(\pi T[\delta + i (t\pm x)]\right)}
\right]^g\ .
\end{equation}   
Using this transformation, we can recalculate the ${\tilde P_g}$'s and
obtain their finite $T$ version:
\begin{eqnarray}
{\tilde P_g(\omega,x,T)}&=&
\int_{-\infty}^\infty dt\ e^{i\omega t}
\left[\frac{\sin\left(\pi T[\delta + i(t+x)]\right)}{\pi T}
\right]^{-g}
\left[\frac{\sin\left(\pi T[\delta + i(t-x)]\right)}{\pi T}
\right]^{-g}\\
&=&(\pi T)^{2g}
\int_{-\infty}^\infty dt\ e^{i\omega t}
\ e^{-i\frac{\pi}{2}g[{\rm sign}(t-x)+{\rm sign}(t+x)]}
\bigg[\sinh (\pi T|t-x|) \sinh (\pi T|t+x|)\bigg]^{-g}\ .
\end{eqnarray}
What we need for the calculation of the currents is the difference
${\tilde P_g(\omega,x,T)}-{\tilde P_g(-\omega,x,T)}$, which
simplifies to
\begin{equation}
{\tilde P_g(\omega,x,T)}-{\tilde P_g(-\omega,x,T)}=
4(\pi T)^{2g}\ \sin(\pi g)\ {\rm Im}\bigg\{
\int_{|x|}^\infty dt\ e^{- i\omega t}
\bigg[\sinh (\pi T|t-x|) \sinh (\pi T|t+x|)\bigg]^{-g}\bigg\}\ .
\end{equation}
After calculating the integral above, we find that it can be written
as
\begin{equation}
{\tilde P_g(\omega,x,T)}-{\tilde P_g(-\omega,x,T)}=
\left[{\tilde P_g(\omega,x=0,T)}-{\tilde P_g(-\omega,x=0,T)}\right]\ 
\times\ H_g(\omega,x,T)\ ,
\end{equation}
where the $x=0$ difference is
\begin{equation}
{\tilde P_g(\omega,x=0,T)}-{\tilde P_g(\omega,x=0,T)}=
4(\pi T)^{2g-1}\ 
B\left(g-i{\omega \over 2\pi T},g+i{\omega \over 2\pi T}\right)
\cosh \left({\omega \over 2 T}\right)\ ,
\end{equation}
and the scaling factor $H_g(\omega,x,T)$ for $x\ne 0$ is
\begin{equation}
H_g(\omega,x,T)=2\pi{\Gamma(2g) \over \Gamma(g)} {e^{-2g\pi T|x|} \over
\sinh {\omega \over 2 T}}\
\times\ {\rm Im}
\bigg\{
{e^{i\omega |x|}\ F(g,g-i{\omega \over 2\pi T};
1-i{\omega \over 2\pi T};e^{-4\pi T|x|})
\over
\Gamma(g+i{\omega \over 2\pi T})\ \Gamma(1-i{\omega \over 2\pi T})}
\bigg\}\ ,
\end{equation}   
where $F$ is the hypergeometric function.

Again, the tunneling current can be written as the tunneling through a
single contact:
\begin{equation}
I_t=e^*
|\Gamma_{\rm eff}|^2 \ 
4(\pi T)^{2g-1}\ 
B\left(g-i{\omega \over 2\pi T},g+i{\omega \over 2\pi T}\right)
\sinh \left({\omega \over 2 T}\right)\ ,
\end{equation}
but with an effective coupling
\begin{equation}
|\Gamma_{\rm eff}|^2=\sum_{i,j=1}^N \Gamma_i\ \Gamma^*_j\ H_g(\omega_J,
|x_i-x_j|,T)\ ,
\end{equation}
much in the same way as in the $T=0$ case.


%
%
%
%
%
%


\begin{thebibliography}{99}
\bibitem{footnote2} Here we are clearly assuming that the edge modes have a
linear dispersion on an energy scale on which there is significant
variation in the transmission; otherwise the lineshape with $V$ will be
distorted.
\bibitem{footnote3} Alternatively, one can account for Aharonov-Bohm and
fractional statistics oscillations within the scope of edge state
theory \cite{XGWcll}; they can be absorbed into the relative phase
between the two tunneling amplitudes $\Gamma_1$ and $\Gamma_2$ across
the contacts of the interferometer. It is thus sheer question of taste
as to which framework one chooses to use: that which focuses on the
bulk, or that which looks at the boundary.
\bibitem{footnote4} In writing this form, 
we have assumed that there is no significant energy dependence of the 
quasiparticle tunneling at the (low) energies of interest. This is
problematic only in our analysis of the Fabry-Perot oscillations where
a range of energies enter and the absolute location of the (AB) nodes is
sensitive to this assumption. While this will require a quantitative
assesment for realistic devices, we should note that regardless, a 
comparision between integer and fractional states could well serve
as a diagnostic of Luttinger liquid behavior in the latter.
\end{thebibliography}
\end{document}